\documentclass{elsart}
\newtheorem{ex}[thm]{Example}
\usepackage{amssymb}
\begin{document}
\begin{frontmatter}


%

\title{
$\!\!\!\!$
Continuous and Discrete Homotopy Operators: 
A Theoretical Approach made Concrete\thanksref{nsf}}
\thanks[nsf]{This material is based upon work supported by the 
National Science Foundation under Grants Nos.\ CCR-9901929,
DMS-9732069, DMS-9912293, and DMS-FRG-0351466.}
%
%
\author[hereman]{W.\ Hereman\corauthref{corres}},
\corauth[corres]{Corresponding author.}
\ead{whereman@mines.edu}
\ead[url]{www.mines.edu/fs$\_$home/whereman}
\author[deconinck]{B.\ Deconinck},
\ead{bernard@amath.washington.edu}
\author[hereman]{L.\ D.\ Poole}
\ead{lpoole@mines.edu}
\address[hereman]{Department of Mathematical and Computer Sciences,
         Colorado School of Mines,
         Golden, CO 80401-1887, USA.}
\address[deconinck]{Department of Applied Mathematics,
           University of Washington,
           Seattle, WA 98195-2420, USA.}
\begin{abstract}
%
Using standard calculus, explicit formulas for the one-dimensional 
continuous and discrete homotopy operators are derived.
It is shown that these formulas are equivalent to those in terms of 
Euler operators obtained from the variational complex.

The continuous homotopy operator automates integration by parts on the 
jet space. 
Its discrete analogue can be used in applications where summation by 
parts is essential. 
Several example illustrate the use of the homotopy operators.

The calculus-based formulas for the homotopy operators are easy to implement 
in computer algebra systems such as {\it Mathematica} and {\it Maple}.
The homotopy operators can be readily applied to the symbolic computation of 
conservation laws of nonlinear partial differential equations and 
differential-difference equations. 
%
%
%
%
%
\end{abstract}
%
\begin{keyword}
%
%
homotopy operator 
\sep exactness 
\sep conservation law 
\sep computer algebra
%
\PACS 
02.30.Ik \sep 05.45.Yv \sep 02.30.Jr \sep 11.30.-j \sep 02.70.Wz
\end{keyword}
\end{frontmatter}
%
%
\vspace{-15mm}
\section{Introduction}
\label{sectionintroduction}
\vspace{-8mm}
%
During the development of symbolic algorithms 
\cite{WHquanchem06,Heremanetalbookchapter05,WHetalcrm04} 
for the computation of conservation laws of nonlinear partial differential 
equations (PDEs) and nonlinear differential-difference equations (DDEs) 
we encountered powerful tools from the calculus of variations and 
differential geometry that deserve attention in their own right.
This paper focuses on one of these tools: the homotopy operator. 

Inspired by work of Kruskal {\it et al.} \cite{MKetal70}, we give a 
straightforward derivation of formulas for the continuous homotopy operator 
and its discrete counterpart.
We show that our formulas are equivalent to those in terms of Euler 
operators obtained from the variational complex \cite{WHetalcrm04}.
For lack of space, we only cover the one-dimensional cases (1D) with 
independent variable $x$ or lattice variable $n.$ 
The generalization to multiple independent variables is cumbersome 
\cite{Heremanetalbookchapter05}. 


The continuous homotopy operator goes back to Volterra's work \cite{VV13} 
on the inverse problem of the calculus of variations. 
The homotopy operator also appears in the proof of the converse of 
Poincar\'e's lemma \cite{PObook93}, which states that exact differential 
forms are closed and vice versa (at least on a star-shaped domain in 
Euclidean space). 
Poincar\'e's lemma is a special case of the so-called {\it de Rham complex}, 
where one investigates the equivalence of closedness and exactness of 
differential $k-$forms in generality. 
The key to exactness proofs of various complexes (such as the de Rham complex) 
is the construction of suitable homotopy operators \cite{PObook93}. 

In basic terms, the 1D continuous homotopy operator reduces the problem 
of integration by parts on the jet space to a sequence of differentiations 
followed by a single definite integration with respect to an auxiliary 
variable. 
In 2D and 3D, the homotopy operator allows one to invert the total divergence 
operator \cite{PObook93}.
Irrespective of the number of independent or dependent variables, the problem 
can be reduced to a single definite integral. 
Applications of the continuous homotopy operator in multi-dimensions 
can be found in \cite{WHquanchem06,Heremanetalbookchapter05,PObook93}.

Likewise, the discrete homotopy operator is a tool to invert the 
forward difference operator whatever the application is.
It circumvents summation by parts by applying shifts and differentiations 
followed by a one-dimensional integration with respect to an auxiliary 
variable \cite{WHetalcrm04}.
Applications of the discrete homotopy operator are given in 
\cite{Heremanetalbookchapter05,WHetalcrm04}.

As shown in \cite{PHandEMfcm04,EMandPHams02}, the parallelism between the 
continuous and discrete cases can be made rigorous as both can be formulated 
in terms of variational bicomplexes.
We do not use the abstract framework in order to make this paper accessible 
to as wide an audience as possible.
Aficionados of {\it de Rham} complexes should consult \cite{IA92,IAbook04}
and \cite{PHandEMfcm04,EMandPHams02,EMandRQ04}.
The latter set of papers covers the discrete variational bicomplex.

Several examples illustrate the inner workings of the homotopy operator at 
the calculus level, without ``wedges and hooks" or differential forms.
Avoiding sophisticated arguments from differential geometry, we can 
introduce the powerful concept of homotopy operators to a wider audience. 

%
%
In \cite{WHquanchem06,Heremanetalbookchapter05,WHetalcrm04} we apply 
homotopy operators to the symbolic computation of conservation laws of 
nonlinear PDEs and DDEs.
Beyond DDEs, the discrete homotopy operator is useful in 
the study of difference equations \cite{PHandEMfcm04,EMandPHams02,EMandRQ04}.
%
%

Despite their universality and applicability, homotopy operators have not
been implemented in major computer algebra systems (CAS) like 
{\it Mathematica} and {\it Maple}. 
CAS offer few reliable tools for integration (or summation) by parts of 
expressions involving {\it unknown} functions and their derivatives 
(or shifts). 
%
We hope that the calculus-based formulas for the homotopy operator presented 
in this paper will lead to more sophisticated integration algorithms within 
CAS.

In summary, our paper has the following objectives: 
(i) Give a straightforward derivation of the 1D continuous homotopy operator, 
(ii) Present the discrete homotopy operator by analogy with the continuous 
case, 
(iii) Illustrate the inner workings of the homotopy operators at the calculus 
level,  
(iv) Present alternate, readily applicable formulas for the homotopy 
operators which lead to efficient and fast symbolic codes for integration 
and summation by parts.
\vspace{-8mm}
\section{Derivation of the continuous homotopy operator in 1D}
\label{sectioncontinuoushomotopyoperator}
\vspace{-8mm}
In \cite{WHquanchem06,Heremanetalbookchapter05}, 
we presented the homotopy operator and referred to \cite{PObook93} 
for a proof, which involves working with differential forms.
Inspired by \cite{MKetal70}, we give a calculus-based derivation of the 
homotopy operator ${\mathcal{H}}_{u(x)}.$
For simplicity and clarity, we show the derivation for one dependent variable 
$u$ and one independent variable $x$ (henceforth referred to as the 1D case).

To do so, we first introduce a ``degree" operator ${\mathcal{M}}$ and 
its inverse, the total derivative ${\mathcal{D}}_x,$ and the Euler 
operator ${\mathcal{L}}^{(0)}_{u(x)}.$ 
The latter is also called the variational derivative or Euler-Lagrange 
operator.

The calculations below are carried out in the {\it jet space} where one 
treats $u, u_x, u_{2x},$ etc., as independent.
As usual, $u_x = \frac{\partial u}{\partial x}, 
u_{2x} = \frac{\partial^2 u}{\partial x^2}$, etc.\
The operators act on $f(u, u_x, u_{2x},\ldots, u_{Mx}).$ 
%
%
Such functions are called {\it differential functions} \cite{PObook93}.
Throughout this paper we assume that the differential functions lack 
constant terms and that the upper bounds in the summations equal the 
order $M$ of the differential function the operators are applied to.
%
\vspace{-2mm}
\begin{defn}
The degree operator ${\mathcal{M}}$ is defined by 
\vspace{-2mm}
\begin{equation}
\label{operM}
{\mathcal{M}} f 
= \sum_{i=0}^{M} u_{ix} \frac{\partial f}{\partial u_{ix}} 
= u \frac{\partial f}{\partial u} 
+ u_x \frac{\partial f}{\partial u_{x}} 
+ u_{2x} \frac{\partial f}{\partial u_{2x}} 
+ \cdots 
+ u_{Mx} \frac{\partial f}{\partial u_{Mx}}, 
\end{equation}
\vspace{-7mm}
where $f$ is a differential function of order $M.$
\end{defn}
\vspace{-2mm}
\begin{ex}
If $f = u^p u_x^q u_{3x}^r,$ where $p,q,$ and $r$ are non-negative 
integers, then
\vspace{-3mm}
\begin{equation}
\label{applofM}
g = {\mathcal{M}} f
= \sum_{i=0}^{3} u_{ix} \frac{ \partial f}{\partial u_{ix}} 
= (p + q + r) \, u^p u_x^q u_{3x}^r.
\end{equation}
\vspace{-5mm}
\end{ex}
Thus, application of ${\mathcal{M}}$ to a monomial results in multiplication 
of the monomial with its {\it degree}, i.e.\ the total number of factors 
in that monomial.

We use the homotopy concept to construct the inverse operator, 
${\mathcal{M}}^{-1}.$ 
Given a differential function $g(u),$ let $g[\lambda u]$ denote $g(u)$ 
where $u$ is replaced by $\lambda u,$ $u_x$ is replaced by $\lambda u_x,$
etc., where $\lambda$ is an auxiliary parameter.
We now show 
\vspace{-2mm}
\begin{equation}
\label{inverseM}
{\mathcal{M}}^{-1} g(u)
= \int_{0}^{1} g[\lambda u] \,\frac{d \lambda}{\lambda}.
\end{equation}
\vspace{-7mm}
Indeed, if $g(u)$ has order $M$ then so does $g[\lambda u],$ and
\vspace{-2mm}
\begin{equation}
\label{derivminverse1}
\frac{d}{d \lambda} g[\lambda u]
= \sum_{i=0}^{M} \frac{\partial g[\lambda u]}{\partial \lambda u_{ix}}
\frac{d \lambda u_{ix}}{d \lambda} 
= \frac{1}{\lambda} \sum_{i=0}^{M} u_{ix} 
\frac{\partial g[\lambda u]}{\partial u_{ix}} 
= \frac{1}{\lambda} {\mathcal{M}} g[\lambda u].
\end{equation}
\vspace{-7mm}
Upon integration of both sides with respect to $\lambda,$ we get
\vspace{-2mm}
\begin{equation}
\label{derivminverse2}
\!\!\!\!\!\!\!\!\!\!
\int_{0}^{1} \frac{d}{d \lambda} g[\lambda u] \, d \lambda 
= \left. g[\lambda u]\right|_{\lambda = 0}^{\lambda = 1} 
\!=\! g(u) - g(0) 
= \int_{0}^{1} {\mathcal{M}}   g[\lambda u] \, \frac{d \lambda}{\lambda}
= {\mathcal{M}} \!\int_{0}^{1} g[\lambda u] \, \frac{d \lambda}{\lambda} .
\end{equation}
\vspace{-7mm}
Assuming $g(0) = 0$ and applying ${\mathcal{M}}^{-1}$ to both sides
of (\ref{derivminverse2}), Eq.\ (\ref{inverseM}) readily follows.
The assumption $g(0) = 0$ restricts the choice for $f.$
We only consider differential functions involving monomials in 
$u, u_x,$ etc., and on occasion multiplied by $\sin u$ or $\cos u.$
%
\noindent
\vspace{-2mm}
\begin{ex}
For $g$ in (\ref{applofM}), 
we have 
$g[\lambda u] = (p + q + r) \lambda^{p + q + r} \, u^p u_x^q u_{3x}^r.$
\newline
Using (\ref{inverseM}), 
\vspace{-5mm}
\begin{eqnarray}
\label{appofMinv}
{\mathcal{M}}^{-1} g
&=& \int_{0}^{1} (p + q + r) \, {\lambda}^{p + q + r - 1} \,
u^p u_x^q u_{3x}^r \; d \lambda \\
&=&  u^p u_x^q u_{3x}^r 
\, \left. {\lambda}^{p + q + r} \right|_{\lambda = 0}^{\lambda = 1} 
= u^p u_x^q u_{3x}^r.
\end{eqnarray}
\vspace{-2mm}
\end{ex}
\vspace{-2mm}
\begin{defn}
The total derivative operator ${\mathcal{D}}_x$ is defined by 
\vspace{-2mm}
\begin{equation}
\label{operD}
\!\!\!\!\!\!
{\mathcal{D}}_x f = 
\sum_{i=0}^{M} u_{(i+1)x} \frac{\partial f}{\partial u_{ix}} 
= u_x \frac{\partial f}{\partial u} 
+ u_{2x} \frac{\partial f}{\partial u_{x}} 
+ \cdots 
+ u_{(M+1)x} \frac{\partial f}{\partial u_{Mx}}.
\end{equation}
\vspace{-7mm}
\end{defn}
\noindent
\vspace{-2mm}
\begin{ex}
If $f = u^p u_x^q u_{3x}^r,$ then
\vspace{-2mm}
\begin{equation}
\!\!\!\!\!\!
{\mathcal{D}}_x f = 
\sum_{i=0}^{3} u_{(i+1)x} \frac{\partial f}{\partial u_{ix}}
= p u^{p-1} u_x^{q} u_{3x}^r 
+ q u^{p}   u_x^{q-1} u_{2x} u_{3x}^r 
+ r u^{p}   u_x^{q} u_{3x}^{r-1} u_{4x}.
\end{equation}
\vspace{-7mm}
\end{ex}
\noindent
\vspace{-2mm}
\begin{thm}
\label{commutation}
The operators ${\mathcal{M}}$ and ${\mathcal{D}}_x$ commute as do 
${\mathcal{M}}^{-1}$ and ${\mathcal{D}}_x.$
\end{thm}
\noindent
{\it Proof.} The proof that ${\mathcal{M}}$ and ${\mathcal{D}}_x$ commute 
is straightforward: applying ${\mathcal{M}}$ to ${\mathcal{D}}_x f$ gives 
the same result as applying ${\mathcal{D}}_x$ to ${\mathcal{M}} f,$ 
using standard calculus manipulations.
Proving that ${\mathcal{M}}^{-1}$ commutes with ${\mathcal{D}}_x$ 
is then immediate. 
$\blacksquare$
\vspace{-3mm}
\begin{defn}
The continuous Euler operator of order zero (variational derivative) 
${\mathcal{L}}^{(0)}_{u(x)}$ is defined \cite{PObook93} by 
\vspace{-3mm}
\begin{eqnarray}
\label{zeroeulerscalarux}
\!\!\!\!\!\!
\mathcal{L}^{(0)}_{u(x)} f
&=& \sum_{k=0}^{M} (-{\mathcal{D}}_x)^k \frac{\partial f}{\partial u_{kx} } 
\nonumber \\
&=& \frac{\partial f}{\partial {u} } 
- {\mathcal{D}}_x \frac{\partial f}{\partial {u_x} }
+ {\mathcal{D}}_{x}^2 \frac{\partial f}{\partial {u_{2x}} } 
- {\mathcal{D}}_{x}^3 \frac{\partial f}{\partial {u_{3x}} } 
+ \cdots 
+ (-1)^M {\mathcal{D}}_{x}^M \frac{\partial f}{\partial {u_{Mx}} }.
\end{eqnarray}
\vspace{-2mm}
\end{defn}
\vspace{-4mm}
\begin{defn}
A differential function $f$ of order $M$ is called exact if there exists 
a differential function $F$ of order $M-1$ so that $f = {\mathcal{D}}_x F.$
\end{defn}
\vspace{-2mm}
\begin{thm}
A necessary and sufficient condition for a differential function $f$ 
to be exact is that ${\mathcal{L}}_{u(x)}^{(0)} f \equiv 0.$
\end{thm} 
\vspace{-1mm}
\noindent
{\it Proof.} A proof is given in e.g.\ \cite{MKetal70}.
\vspace{-2mm}
\begin{ex}
\label{examplefexact}
Let $f\!\!=\!\!2 u_x u_{2x} \cos u\!-\!u_x^3 \sin u.$
Note that $f\!\!=\!\!{\mathcal{D}}_x F$ with $F\!\!=\!u_x^2 \cos u.$

We show that $f$ is indeed exact. 
Using (\ref{zeroeulerscalarux}), we readily verify that 
\vspace{-6mm}
\begin{eqnarray}
\!\!\!\!\!\!\!
\mathcal{L}^{(0)}_{u(x)} f 
&=&
\frac{\partial f}{\partial u} 
- {\mathcal{D}}_x \frac{\partial f}{\partial u_x}
+ {\mathcal{D}}_{x}^2 \frac{\partial f}{\partial u_{2x}}
\nonumber \\
&=&
-\! 2 u_x u_{2x} \sin u \!-\! u_x^3 \cos u
\!-\! {\mathcal{D}}_x [ 2 u_{2x} \cos u \!-\! 3 u_x^2 \sin u ] 
\!+\! {\mathcal{D}}_x^2 [ 2 u_x \cos u ] 
\nonumber \\
&=&
- 2 u_x u_{2x} \sin u - u_x^3 \cos u
- [ 2 u_{3x} \cos u - 8 u_x u_{2x} \sin u - 3 u_x^3 \cos u ] 
\nonumber \\
&& +\, [ 2 u_{3x} \cos u - 6 u_x u_{2x} \sin u - 2 u_x^3 \cos u ] 
\equiv 0.
\end{eqnarray}
\vspace{-4mm}
\end{ex}
\vspace{-4mm}
\begin{defn}
The continuous homotopy operator with variable $u(x)$ is
\vspace{-2mm}
\begin{equation}
\label{homotopyscalarux}
\mathcal{H}_{u(x)} f = 
\int_{0}^{1} \left( I_{u} f \right) [\lambda u] \, \frac{d \lambda}{\lambda}, 
\end{equation}
\vspace{-7mm}
where the integrand $I_{u} f$ is given by 
\vspace{-2mm}
\begin{equation}
\label{integrandhomotopyscalarux}
I_{u} f 
= \sum_{i=0}^{M-1} u_{ix} \sum_{k=i+1}^M (-{\mathcal{D}}_{x})^{k-(i+1)} 
\frac{\partial f}{\partial u_{kx}}.
\end{equation}
\vspace{-7mm}
\end{defn}
%
%
\noindent
\vspace{-2mm}
\begin{thm}
\label{theoremhomotopyux}
Given an exact differential function $f$ of order $M$ one has 
\newline
$F = {\mathcal{D}}_x^{-1} f = \int f \, dx = \mathcal{H}_{u(x)} f.$
\end{thm}
\noindent
{\it Proof.} We multiply ${\mathcal{L}}^{(0)}_{u(x)} f 
= \sum_{k=0}^{M} (-{\mathcal{D}}_x)^k \frac{\partial f}{\partial u_{kx} }$ 
by $u$ to restore the degree.
Next, we split off $u \frac{\partial f}{\partial u}.$ 
Then, we integrate by parts and split off  
$u_x \frac{\partial f}{\partial u_{x}}.$ 
We repeat this process until we split off 
$u_{Mx} \frac{\partial f}{\partial u_{Mx}}.$
%
%
In detail, 
\vspace{-8mm}
\begin{eqnarray}
\label{proofhomotopyux}
u {\mathcal{L}}^{(0)}_{u(x)} f 
&=& 
u \sum_{k=0}^{M} (-{\mathcal{D}}_x)^k \frac{\partial f}{\partial u_{kx} }
\nonumber \\
&=& 
u \frac{\partial f}{\partial u} 
- {\mathcal{D}}_x 
\left(
u \sum_{k=1}^{M} (-{\mathcal{D}}_x)^{k-1} \frac{\partial f}{\partial u_{kx} } 
\right) 
+ u_x \sum_{k=1}^{M}(-{\mathcal{D}}_x)^{k-1}\frac{\partial f}{\partial u_{kx}} 
\nonumber \\
&=& 
u \frac{\partial f}{\partial u} + u_x \frac{\partial f}{\partial u_{x}} 
- {\mathcal{D}}_x 
\left(
u \sum_{k=1}^{M} (-{\mathcal{D}}_x)^{k-1} \frac{\partial f}{\partial u_{kx} } 
\nonumber \right. \\
&& 
\left. 
+ u_x \sum_{k=2}^{M}(-{\mathcal{D}}_x)^{k-2}\frac{\partial f}{\partial u_{kx}} 
\right) 
+ u_{2x} 
\sum_{k=2}^{M}(-{\mathcal{D}}_x)^{k-2}\frac{\partial f}{\partial u_{kx}} 
\nonumber \\
&=& \ldots 
\nonumber \\
&=& 
u \frac{\partial f}{\partial u} + u_x \frac{\partial f}{\partial u_{x}} 
+ \ldots + 
u_{Mx} \frac{\partial f}{\partial u_{Mx}} 
- {\mathcal{D}}_x 
\left(
u \sum_{k=1}^{M} (-{\mathcal{D}}_x)^{k-1} \frac{\partial f}{\partial u_{kx} } 
\right. 
\nonumber \\
&& \left. 
+ 
u_x \sum_{k=2}^{M} (-{\mathcal{D}}_x)^{k-2}\frac{\partial f}{\partial u_{kx}
} 
+ \ldots 
+ u_{(M-1)x} \sum_{k=M}^{M} (-{\mathcal{D}}_x)^{k-M}
\frac{\partial f}{\partial u_{kx} } 
\right)
\nonumber \\
&=& \sum_{i=0}^{M} u_{ix} \frac{\partial f}{\partial u_{ix}} 
- {\mathcal{D}}_x 
\left(
\sum_{i=0}^{M-1} u_{ix} \sum_{k=i+1}^{M} (-{\mathcal{D}}_x)^{k-(i+1)}
\frac{\partial f}{\partial u_{kx} }
\right) 
\nonumber \\
&=& {\mathcal{M}} f - {\mathcal{D}}_x 
\left(
\sum_{i=0}^{M-1} u_{ix} \sum_{k=i+1}^{M} (-{\mathcal{D}}_x)^{k-(i+1)}
\frac{\partial f}{\partial u_{kx} }
\right).
\end{eqnarray}
\vspace{-4mm}
Since $f$ is exact we have ${\mathcal{L}}^{(0)}_{u(x)} f = 0.$ 
Eq.\ (\ref{proofhomotopyux}) then implies
\vspace{-2mm}
\begin{equation}
\label{mf}
{\mathcal{M}} f = {\mathcal{D}}_x 
\left(
\sum_{i=0}^{M-1} u_{ix} \sum_{k=i+1}^{M} (-{\mathcal{D}}_x)^{k-(i+1)}
\frac{\partial f}{\partial u_{kx} }
\right).
\end{equation}
\vspace{-7mm}
Applying ${\mathcal{M}}^{-1}$ and using 
${\mathcal{M}}^{-1} {\mathcal{D}}_x = {\mathcal{D}}_x {\mathcal{M}}^{-1}$ 
from Theorem~\ref{commutation}, we obtain
\vspace{-2mm}
\begin{equation}
\label{expressionf}
f = {\mathcal{D}}_x \left(
{\mathcal{M}}^{-1}
\sum_{i=0}^{M-1} u_{ix} \sum_{k=i+1}^{M} (-{\mathcal{D}}_x)^{k-(i+1)}
\frac{\partial f}{\partial u_{kx} } 
\right). 
\end{equation}
\vspace{-7mm}
Applying ${\mathcal{D}}_x^{-1}$ 
and using (\ref{inverseM}), we get 
\vspace{-2mm}
\begin{equation}
\label{expressionF}
\!\!\!\!\!\!\!\!
F = {\mathcal{D}}_x^{-1} f =
\int_{0}^{1} \left(
\sum_{i=0}^{M-1} u_{ix} \sum_{k=i+1}^{M} (-{\mathcal{D}}_x)^{k-(i+1)} 
\frac{\partial f}{\partial u_{kx} } \right) [\lambda u] 
\, \frac{d \lambda}{\lambda} 
= \mathcal{H}_{u(x)} f 
\end{equation}
\vspace{-7mm}
using (\ref{homotopyscalarux}) and (\ref{integrandhomotopyscalarux}).
$\blacksquare$
\vspace{-2mm}
\begin{ex}
Let $f = 2 u_x u_{2x} \cos u - u_x^3 \sin u$ which is exact as shown in 
Example~\ref{examplefexact}.
Using (\ref{integrandhomotopyscalarux}) with $M=2,$ we readily compute
\vspace{-5mm}
\begin{eqnarray} 
I_{u} f &=& 
\sum_{i=0}^{1} u_{ix} 
\sum_{k=i+1}^{2} (-{\mathcal{D}}_x)^{k-i-1} \frac{\partial f}{\partial u_{kx}} 
= u \frac{\partial f}{\partial u_{x}} 
- u {\mathcal{D}}_x (\frac{\partial f}{\partial u_{2x}})
+ u_x (\frac{\partial f}{\partial u_{2x}})
\nonumber \\
&=& u ( 2 u_{2x} \cos u - 3 u_x^2 \sin u ) 
- u {\mathcal{D}}_x (2 u_x \cos u)
+ u_x (2 u_x \cos u)
\nonumber \\
&=& - u u_x^2 \sin u + 2 u_x^2 \cos u.
\end{eqnarray} 
\vspace{-2mm}
\noindent 
Using (\ref{homotopyscalarux}), 
%
\vspace{-5mm}
\begin{eqnarray}
F &=& \mathcal{H}_{u(x)} f 
= \int_0^1 \left( I_{u} f \right) [\lambda u] \, \frac{d\lambda}{\lambda} 
= \int_0^1 \left( -\lambda^2 u u_x^2 \sin(\lambda u) 
+ 2 \lambda u_x^2 \cos(\lambda u) \right) \, d\lambda  
\nonumber \\ 
&=& u_x^2 \cos u.
\end{eqnarray}
\end{ex}
\vspace{-5mm}
\noindent
Thus, application of (\ref{homotopyscalarux}) yields $F$ without integration 
by parts with respect to $x.$
Indeed, $F$ can be computed via repeated differentiation followed by a 
one-dimensional integration with respect to an auxiliary variable $\lambda.$
%
%
\vspace{-8mm}
\section{Alternate form of the continuous homotopy operator in 1D}
\label{sectionalternatecontinuoushomotopyoperator}
\vspace{-8mm}
Formulas (\ref{homotopyscalarux}) and 
(\ref{integrandhomotopyscalarux})
are valid for one dependent variable $u.$ 
In \cite{Heremanetalbookchapter05} we presented a calculus-based formula 
for the homotopy operator for $N$ dependent variables in 1D based on work 
by Anderson and Olver in \cite[p.\ 372]{PObook93}:   
\vspace{-2mm}
\begin{equation}
\label{homotopyvectorux}
\mathcal{H}_{{\bf u}(x)} f = 
\int_{0}^{1} \sum_{j=1}^{N} \left( I_{u^{(j)}} f \right) [\lambda {\bf u}] 
\, \frac{d \lambda}{\lambda},
\end{equation}
\vspace{-7mm}
where $u^{(j)}$ is the $j$th component of 
${\bf u} = ( u^{(1)}, u^{(2)}, \cdots, u^{(j)}, \cdots, u^{(N)} ).$ 
The integrand,
\vspace{-2mm}
\begin{equation}
\label{integrandhomotopyvectorux}
I_{u^{(j)}} f = \sum_{i=0}^{M^{(j)}-1} {\mathcal{D}}_{x}^i 
\left( u^{(j)} \, {\mathcal{L}}^{(i+1)}_{u^{(j)}(x)} f \right), 
\end{equation}
\vspace{-7mm}
where $M^{(j)}$ is the order of the variable $u^{(j)}$ in $f,$
involves the continuous 1D higher Euler operators \cite{MKetal70,PObook93} 
defined as follows. 
\vspace{-2mm}
\begin{defn}
The continuous higher Euler operators for component $u^{(j)}(x)$ are
%
\vspace{-2mm}
\begin{equation}
\label{highereulervectorux}
{\mathcal{L}}^{(i)}_{u^{(j)}(x)} f = 
\sum_{k=i}^{M^{(j)}} {k \choose i} (-{\mathcal{D}}_x)^{k-i} 
\frac{\partial f}{\partial {u^{(j)}}_{kx}} , 
\end{equation}
\vspace{-7mm}
where ${k \choose i}$ is the binomial coefficient. 
\end{defn}
\noindent 
%
%
Note that the higher Euler operator for $i=0$ and one dependent variable 
$u^{(1)}(x) = u(x)$ matches the variational derivative 
(\ref{zeroeulerscalarux}).
%

In the case of one dependent variable, $u,$ we denote $M^{(1)}$ by $M.$ 
After substitution of ${\mathcal{L}}^{(i+1)}_{u(x)} f$ into 
(\ref{integrandhomotopyvectorux}), we obtain
%
\vspace{-2mm}
\begin{equation}
\label{alternateintegrandhomotopyscalarux}
I_{u} f 
= \sum_{i=0}^{M-1} {\mathcal{D}}_{x}^i 
\left( u \sum_{k=i+1}^{M} {k \choose {i+1}} (-{\mathcal{D}}_x)^{k-(i+1)} 
\frac{\partial f}{\partial u_{kx}} \right) .
\end{equation}
\vspace{-7mm}
\noindent
\vspace{-2mm}
\begin{thm}
\label{equalitycontinuousalternate}
The integrands (\ref{integrandhomotopyscalarux}) and 
(\ref{alternateintegrandhomotopyscalarux}) are equal. 
\end{thm}
\noindent
{\it Proof.} Starting from (\ref{alternateintegrandhomotopyscalarux}), 
we use Leibniz's rule to propagate ${\mathcal{D}}_{x}$ to the right 
\vspace{-5mm}
\begin{eqnarray}
\label{proofequalitycontinuousalternatepart1}
I_{u} f &=& 
\sum_{i=0}^{M-1} {\mathcal{D}}_{x}^i \left(  
u \sum_{k=i+1}^{M} {k \choose {i+1}} (-{\mathcal{D}}_x)^{k-(i+1)} 
\frac{\partial f}{\partial u_{kx}} \right) 
\nonumber \\
&=&
\sum_{i=0}^{M-1} \sum_{j=0}^{i} {i \choose j} u_{jx}
{\mathcal{D}}_{x}^{i-j} \left(  
\sum_{k=i+1}^{M} {k \choose {i+1}} (-{\mathcal{D}}_x)^{k-(i+1)} 
\frac{\partial f}{\partial u_{kx}} \right) 
\nonumber \\
&=& 
\sum_{i=0}^{M-1} \sum_{j=0}^{i} {i \choose j} u_{jx} (-1)^{j-i}
\sum_{k=i+1}^{M} {k \choose {i+1}} {(-\mathcal{D}}_{x})^{k-(j+1)} 
\frac{\partial f}{\partial u_{kx}} .
\end{eqnarray}
Next, we interchange the sums over $i$ and $j$ (to bring $u_{jx}$ up front),
followed by an interchange of the sums over $i$ and $k$ (to bring
${\mathcal{D}}_x$ and 
${\partial f}/{\partial u_{kx}}$ outside the sum over $i).$ 
So, 
\vspace{-7mm}
\begin{eqnarray}
\label{proofequalitycontinuousalternatepart2}
\nonumber \\
I_{u} f
&=&
\sum_{j=0}^{M-1} u_{jx} \sum_{i=j}^{M-1} {i \choose j} (-1)^{j-i}
\sum_{k=i+1}^{M}  {k \choose {i+1}} {(-\mathcal{D}}_{x})^{k-(j+1)} 
\frac{\partial f}{\partial u_{kx}} 
\nonumber \\
&=& 
\sum_{j=0}^{M-1} u_{jx} \sum_{k=j+1}^{M}
{(-\mathcal{D}}_{x})^{k-(j+1)} \frac{\partial f}{\partial u_{kx}} 
\sum_{i=j}^{k-1} (-1)^{i-j} {i \choose j} {k \choose {i+1}} 
\nonumber \\
&=& 
\sum_{j=0}^{M-1} u_{jx} \sum_{k=j+1}^{M}
{(-\mathcal{D}}_{x})^{k-(j+1)} \frac{\partial f}{\partial u_{kx}}, 
\end{eqnarray}
where we have used the identity 
\vspace{-2mm}
\begin{equation}
\label{binidentity1}
\sum_{i=j}^{k-1} (-1)^{i-j} {i \choose j} {k \choose {i+1}} = 1 \quad\;
{\rm if} \; k \ge j + 1, 
\end{equation}
\vspace{-7mm}
which is straightforward to prove using mathematical induction.
$\blacksquare$
%
\vskip 0.001pt
\indent
Homotopy operator (\ref{homotopyvectorux}) can easily be implemented in CAS. 
In our experience, integrand (\ref{integrandhomotopyscalarux}) leads to a
more efficient and faster algorithm for it requires substantially less 
differentiations then (\ref{alternateintegrandhomotopyscalarux}).
\vspace{-8mm}
\section{Discrete Euler and homotopy operators}
\label{sectiondiscretehomotopyoperator}
\vspace{-8mm}
We now turn to the discrete analogues of differential functions, 
Euler operators and homotopy operators. 
For simplicity we consider the case of one lattice variable $n$ which 
results, for example, from a discretization of the variable $x.$ 
We allow $N$ dependent variables, i.e.\
${\bf u}_n = (u^{(1)}_n, u^{(2)}_n, \cdots, u^{(j)}_n, \cdots, u^{(N)}_n ). $ 
For simplicity, in the examples we denote these components by 
$u_n, v_n,$ etc.\ 

By analogy with ${\mathcal{D}}_x$ and ${\mathcal{D}}_x^{-1},$ 
we define shift operators acting on $f_n({\bf u}_n).$ 
%
%
\vspace{-2mm}
\begin{defn} 
${\rm D}$ is the up-shift operator (also known as forward- or right-shift) 
such that ${\rm D} \, f_n = f_{n+1}.$ 
Its inverse, ${\rm D}^{-1},$ is the down-shift operator
(or backward- or left-shift) such that ${\rm D^{-1}} \, f_n = f_{n-1}.$ 
The identity operator is denoted by ${\rm I}.$ 
Lastly, $\Delta = {\rm D} - {\rm I}$ is the forward difference operator 
so that $\Delta \, f_n = ({\rm D} - {\rm I}) \, f_n = f_{n+1} - f_n.$
\end{defn}
Given are functions $f_n$ in discrete variables $u_n, v_n, \ldots $ 
and their up and down shifts.
If $ f_n( u_{n-p}, v_{n-r}, \cdots, u_n, v_n, \cdots, u_{n+q}, v_{n+s} )$ 
with $p \ge r$ involves negative shifts, one must first remove these by 
replacing $f_n$ by ${\tilde f}_n = {\rm D}^p f_n.$
From this point we assume that all negative shifts have been removed.
\vspace{-2mm}
\begin{defn}
$f_n$ is called exact if it is a total difference, 
i.e.\ there exists a $F_n$ so that $f_n = {\rm \Delta} \, F_n.$
\end{defn}
%
\begin{ex}
\label{fdiscreteexact}
Let
%
\begin{equation}
\label{fdiscrete}
f_n = \sin(u_{n+3}) \cos^2(v_{n+2}^2) - \sin(u_{n+1}) \cos^2(v_{n}^2).
%
%
\end{equation}
\vspace{-5mm}
By hand, we readily verify that $f_n = \Delta F_n$ with
\begin{equation}
\label{Fn}
F_n = \sin(u_{n+2}) \cos^2(v_{n+1}^2) + \sin(u_{n+1}) \cos^2(v_{n}^2).
\end{equation}
\vspace{-5mm}
%
%
%
So, $f_n$ is exact.
\end{ex}
\vskip 0.001pt
\noindent
Below we address the following questions.
(i) How can one test whether or not $f_n$ is exact?
Equivalently, how does one know that $F_n$ exists in closed form?
(ii) Can one compute $F_n \!=\! \Delta^{-1} f_n$ in a way analogous to the 
continuous case?
\vspace{-4mm}
\begin{defn}
${\mathcal{L}}^{(0)}_{u^{(j)}_n}$ is the discrete Euler operator of order 
zero (discrete variational derivative) for component $u^{(j)}_n,$ defined 
\cite{VAetaltmp00} by
\vspace{-5mm}
\begin{eqnarray}
\label{eulervectorun}
{\mathcal{L}}^{(0)}_{u^{(j)}_n} f_n &=&
\sum_{k=0}^{M^{(j)}} {{\rm D}}^{-k} 
\frac{\partial f_n}{\partial u^{(j)}_{n+k} } 
= \frac{\partial }{\partial u^{(j)}_n} 
\sum_{k=0}^{M^{(j)}} {\rm D}^{-k} f_n
\nonumber \\
&=& \frac{\partial }{\partial  u^{(j)}_n } 
\left( 
{\rm I} + {\rm D}^{-1} + {\rm D}^{-2} + \cdots + {\rm D}^{-{M^{(j)}}} 
\right) f_n, 
\end{eqnarray}
where $M^{(j)}$ is the highest shift of $u^{(j)}_n$ occurring in $f_n.$ 
\vspace{-2mm}
\end{defn}
%
With respect to the existence of $F_n,$  the following exactness 
criterion is well-known and frequently used \cite{VAetaltmp00}. 
\vspace{-2mm}
\begin{thm}
A necessary and sufficient condition for a function $f_n$ with positive shifts 
to be exact is that ${\mathcal{L}}^{(0)}_{u^{(j)}_n} f_n \equiv 0, 
\; j = 1, 2, 3, \cdots, N.$
\end{thm}
\noindent
{\it Proof.} A proof is given in e.g.\ \cite{MHandWHprsa03}.
$\blacksquare$
\vspace{-3mm}
\begin{ex}
To test that (\ref{fdiscrete}) is exact we apply (\ref{eulervectorun}) 
to $f_n$ for each component of ${\bf u}_n = (u^{(1)}_n, u^{(2)}_n) = 
(u_n,v_n)$ separately.
%
%
For component $u_n$ with maximum shift $M^{(1)} = 3$ we readily verify that
$\mathcal{L}^{(0)}_{u_n} f_n =
\frac{\partial }{\partial{u_n}} 
\left( {\rm I} + {\rm D}^{-1} + {\rm D}^{-2} + {\rm D}^{-3} \right) f_n 
\equiv 0. $
Similarly, for component $v_n$ with maximum shift $M^{(2)} = 2$ we check that
$ \mathcal{L}^{(0)}_{v_n} f_n 
= \frac{\partial }{\partial{v_n}} 
\left( {\rm I} + {\rm D}^{-1} + {\rm D}^{-2} \right) f_n
\equiv 0. $  
\end{ex} 
Next, we compute $F_n$ so that $f_n = {\rm \Delta} \, F_n = F_{n+1} - F_n.$ 
\vspace{-2mm}
\begin{defn}
The discrete homotopy operator for ${\bf u}_n$ is
\vspace{-2mm}
\begin{equation}
\label{homotopyvectorun}
{\mathcal{H}}_{{\bf u}_n} f_n = 
\int_{0}^{1} \sum_{j=1}^{N} 
\left( I_{u^{(j)}_n} f_n \right) [\lambda {\bf u}_n] 
\, \frac{d \lambda}{\lambda},
\end{equation}
\vspace{-7mm}
where the integrand $I_{u^{(j)}_n} f_n$ is given by 
\vspace{-2mm}
\begin{equation}
\label{integrandhomotopyvectorun}
I_{u^{(j)}_n} f_n = 
\sum_{i=0}^{M^{(j)}-1} u^{(j)}_{n+i} 
\frac{\partial }{\partial u^{(j)}_{n+i}}
\sum_{k=i+1}^{M^{(j)}} {{\rm D}}^{-(k-i)} f_n.
%
\end{equation}
\vspace{-7mm}
\end{defn}
\noindent 
As in the continuous case, 
$\left( I_{u^{(j)}_n} f_n \right) [\lambda {\bf u}_n]$ means that after 
$I_{u^{(j)}_n} f_n$ is computed 
\vskip 0.00001pt
\noindent
one replaces 
${\bf u}_n$ by $\lambda {\bf u}_n,$ ${\bf u}_{n+1}$ by 
$\lambda {\bf u}_{n+1},$ etc.

One can use the following theorem 
\cite{WHetalcrm04,PHandEMfcm04,EMandPHams02} to compute $F_n.$ 
\vspace{-2mm}
\begin{thm} 
\label{theoremhomotopyun}
Given an exact function $f_n$ one has 
$F_n = \Delta^{-1} f_n = {\mathcal{H}}_{{\bf u}_n} f_n.$
\end{thm}
\noindent 
{\it Proof.} The proof is similar to that of Theorem~\ref{theoremhomotopyux}. 
For simplicity we show it for one dependent variable, $u^{(1)}_n = u_n,$ 
and we denote $M^{(1)}$ by $M.$
After multiplication of ${\mathcal{L}}^{(0)}_{u_n} f_n 
= \frac{\partial }{\partial u_n} 
\sum_{k=0}^{M} {\rm D}^{-k} f_n$ with $u_n,$
we isolate $u_n \frac{\partial f_n}{\partial u_n}$ and apply 
$\Delta$ to the remaining term in order to sum by parts. 
Next, we isolate $u_{n+1} \frac{\partial f_n}{\partial u_{n+1}},$
followed by another summation by parts. 
\newline
\noindent
This process is repeated $M-1$ times, so that 
\vspace{-3mm}
\begin{eqnarray}
\label{proofhomotopyun}
u_n {\mathcal{L}}^{(0)}_{u_n} f_n 
&=& \sum_{i=0}^{M} u_{n+i} \frac{\partial f_n}{\partial u_{n+i}} 
- {\Delta} \left(
\sum_{i=0}^{M-1} u_{n+i} \frac{\partial }{\partial u_{n+i} }
\sum_{k=i+1}^{M} {\rm D}^{-(k-i)} f_n \right).
\end{eqnarray}
%
The inverse of the discrete operator 
${\mathcal{M}} f_n 
= \sum_{i=0}^{M} u_{n+i} \frac{\partial f_n}{\partial u_{n+i}}$
is computed as ${\mathcal{M}}^{-1} f_n
= \int_{0}^{1} f_n[\lambda u_n] \, \frac{d \lambda}{\lambda}.$
Since $f_n$ is exact
and ${\mathcal{M}}^{-1}$ and $\Delta$ commute, we get
\vspace{-2mm}
\begin{equation}
\label{expressionFn}
\!\!\!\!\!\!\!\!\!\!
F_n = \Delta^{-1} f_n =
\int_{0}^{1} \left( 
\sum_{i=0}^{M-1} u_{n+i} 
\frac{\partial }{\partial u_{n+i}}
\sum_{k=i+1}^M {{\rm D}}^{-(k-i)} f_n \right)\![\lambda u_n] 
\, \frac{d \lambda}{\lambda} 
= {\mathcal{H}}_{u_n} f_n
\end{equation}
\vspace{-7mm}
using (\ref{homotopyvectorun}) and (\ref{integrandhomotopyvectorun}).
$\blacksquare$
\vskip 1pt
\noindent
Thus, the homotopy operator reduces summation by parts needed for the 
inversion of $\Delta$ to a set of shifts and differentiations followed 
by a single definite integral with respect to a scaling parameter $\lambda.$

%
%
\vspace{-2mm}
\begin{ex}
We return to (\ref{fdiscrete}) where $f_n$ involves $u_n$ and $v_n$ with 
maximal shifts $M^{(1)} = 3$ and $M^{(2)} = 2.$
Using (\ref{integrandhomotopyvectorun}), we get
\vspace{-5mm}
\begin{eqnarray}
\!\!\!\!\!\!\!\!\!\!\!\!
I_{u_n} f_n 
&\!=\!&
%
u_n \frac{\partial}{\partial u_n} \!
({\rm D}^{-1} \!+\! {\rm D}^{-2} \!+\! {\rm D}^{-3}) f_n
\!+\! u_{n+1} \frac{\partial}{\partial u_{n+1}} \!
({\rm D}^{-1} \!+\! {\rm D}^{-2}) f_n
\!+\! u_{n+2} \frac{\partial}{\partial u_{n+2}} \!{\rm D}^{-1} f_n 
\nonumber \\
&=& u_n \frac{\partial}{\partial u_n} 
\left( 
\sin(u_{n+2}) \cos^2(v_{n+1}^2) + \sin(u_{n+1}) \cos^2(v_{n}^2) 
\right.
\nonumber \\
&& \left. 
- \sin(u_{n-1}) \cos^2(v_{n-2}^2) - \sin(u_{n-2}) \cos^2(v_{n-3}^2) \right) 
\nonumber \\
&& 
+ u_{n+1} \frac{\partial}{\partial u_{n+1}} 
\left( \sin(u_{n+2}) \cos^2(v_{n+1}^2) - \sin(u_{n}) \cos^2(v_{n-1}^2)  
\right.
\nonumber \\
&& \left. 
+ \sin(u_{n+1}) \cos^2(v_{n}^2) - \sin(u_{n-1}) \cos^2(v_{n-2}^2) \right)
\nonumber \\
&& 
+ u_{n+2} \frac{\partial}{\partial u_{n+2}} 
\left( \sin(u_{n+2}) \cos^2(v_{n+1}^2) - \sin(u_{n}) \cos^2(v_{n-1}^2) 
\right).
\nonumber \\
&=& 
u_{n+1} \cos(u_{n+1}) \cos^2(v_{n}^2) 
+ u_{n+2} \cos(u_{n+2}) \cos^2(v_{n+1}^2).
\end{eqnarray}
and, analogously,
\vspace{-2mm}
\begin{equation}
\!\!\!\!\!\!\!\!\!\!
I_{v_n} f_n 
%
\!=\! -4 \left( v_{n}^2 \sin(u_{n+1}) \cos(v_{n}^2) \sin(v_{n}^2) 
\!+\! v_{n+1}^2 \sin(u_{n+2}) \cos(v_{n+1}^2) \sin(v_{n+1}^2) \right)\!.
\end{equation}
\vspace{-2mm}
\noindent 
Based on (\ref{homotopyvectorun}), we compute 
\vspace{-5mm}
\begin{eqnarray}
F_n &=&
\int_0^1 \left( I_{u_n} f_n + I_{v_n} f_n \right) [\lambda {\bf u}_n]
\, \frac{d\lambda}{\lambda} 
\nonumber \\
&=& 
\int_0^1 \left( u_{n+1} \cos(\lambda u_{n+1}) \cos^2(\lambda^2 v_{n}^2) 
+ u_{n+2} \cos(\lambda u_{n+2}) \cos^2(\lambda^2 v_{n+1}^2) \right.
\nonumber \\
&& \left. \;\;\;\, -\, 4 \, \lambda \, v_{n}^2 \, 
\sin(\lambda u_{n+1}) \cos(\lambda^2 v_{n}^2) \sin(\lambda^2 v_{n}^2) \right.
\nonumber \\
&& \left. \;\;\;\, -\, 4 \, \lambda \, v_{n+1}^2 \, \sin(\lambda u_{n+2}) 
\cos(\lambda^2 v_{n+1}^2) \sin(\lambda^2 v_{n+1}^2) \right) \, d\lambda  
\nonumber \\
&=& \sin(u_{n+2}) \cos^2(v_{n+1}^2) + \sin(u_{n+1}) \cos^2(v_{n}^2) , 
%
%
\end{eqnarray}
which agrees with (\ref{Fn}) previously computed by hand in 
Example~\ref{fdiscreteexact}.
\end{ex}
\vspace{-8mm}
\section{Alternate form of the discrete homotopy operator}
\label{sectionalternatediscretehomotopyoperator}
\vspace{-8mm}
In \cite{Heremanetalbookchapter05} we presented the following formula for 
the discrete homotopy operator with one lattice variable $(n)$ 
and with $N$ dependent variables $u^{(j)}_n$: 
\vspace{-2mm}
\begin{equation}
\label{alternatehomotopyvectorun}
{\mathcal{H}}_{{\bf u}_n} f_n = \int_{0}^{1} \sum_{j=1}^{N} 
\left( I_{u^{(j)}_n} f_n \right) [\lambda {\bf u}_n] 
\, \frac{d \lambda}{\lambda} 
\end{equation}
\vspace{-7mm}
with 
\vspace{-2mm}
\begin{equation}
\label{alternateintegrandhomotopyvectorun}
I_{u^{(j)}_n} f_n = \sum_{i=0}^{M^{(j)}-1} \Delta^i 
    \left( u^{(j)}_n \mathcal{L}^{(i+1)}_{u^{(j)}_n} f_n \right) , 
\end{equation}
\vspace{-7mm}
where the discrete higher Euler operators are defined as follows. 
\vspace{-2mm}
\begin{defn}
The discrete higher Euler operators for component $u^{(j)}_n$ are  
\vspace{-2mm}
\begin{equation}
\label{highereulervectorun}
{\mathcal{L}}^{(i)}_{u^{(j)}_n} f_n
= \sum_{k=i}^{M^{(j)}} {k \choose i}
  {{\rm D}}^{-k} \frac{\partial f_n}{\partial u^{(j)}_{n+k} } 
= \frac{\partial }{\partial u^{(j)}_n } 
  \sum_{k=i}^{M^{(j)}} {k \choose i} {\rm D}^{-k} f_n.
\end{equation}
\vspace{-7mm}
\end{defn}
\noindent
\vspace{-2mm}
\begin{thm}
\label{equalitydiscretealternate}
The integrands (\ref{integrandhomotopyvectorun}) and 
(\ref{alternateintegrandhomotopyvectorun}) are equal, i.e.\
\vspace{-5mm}
\begin{eqnarray}
\label{integrandhomotopyvectorunall}
I_{u^{(j)}_n} f_n 
&=& 
%
%
\sum_{i=0}^{M^{(j)}-1} \Delta^i 
  \left( 
  u^{(j)}_n \frac{\partial }{\partial u^{(j)}_n}
  \sum_{k=i+1}^{M^{(j)}} {k \choose i+1} {{\rm D}}^{-k} f_n 
  \right) 
\nonumber \\
%
%
&=&
\sum_{i=0}^{{M^{(j)}}-1} u^{(j)}_{n+i} 
\frac{\partial }{\partial u^{(j)}_{n+i}} 
\sum_{k=i+1}^{M^{(j)}} {{\rm D}}^{-(k-i)} f_n
%
\end{eqnarray}
\vspace{-5mm}
\end{thm}
\noindent
{\it Proof.} The proof is analogous to that of 
Theorem~\ref{equalitycontinuousalternate}.
$\blacksquare$
\vspace{-8mm}
\section{Conclusions}
\label{conclusions}
\vspace{-8mm}
In this paper we derived formulas for the 1D continuous and discrete 
homotopy operators. 
We showed that our calculus-based formulas are equivalent to those 
obtained from the variational complexes. 
The simplified formulas no longer involve higher Euler operators 
which makes them easier to implement and faster to execute in major 
computer algebra systems.
Simplified versions of the continuous homotopy operator in 2D and 3D 
will be presented elsewhere.
%

%
\end{document}